\documentclass[12pt,english]{article}
\usepackage{ecology}
\usepackage{setspace}
\usepackage{geometry}
\nolinenumbers
\usepackage{fullpage}
\usepackage{graphics}
\usepackage{graphicx}
\usepackage{wrapfig}
\usepackage{tikz}
\usepackage{psfrag}
\usepackage{enumerate}
\usepackage{xspace}
\usepackage{color, soul}
\usepackage{mathrsfs}
\usepackage{amsfonts}
\usepackage{amsmath,amssymb,amsthm}
\usepackage{amstext}
\usepackage{amsopn}
\usepackage{float}
\usepackage{ulem}
\usepackage{multirow}
\usepackage{hyperref}
\usepackage{subcaption}
\usepackage[flushleft]{threeparttable}

\DeclareGraphicsExtensions{.eps, .pdf, .jpg, .jpeg, .tif, .png}

\newcommand{\ds}{\displaystyle}

\begin{document}

\shortauthors{Xu et al.}
\running{Stage-structured blue crab population model}
\articletype{Ecological Applications}
\title{Impacts of density-dependent predation, cannibalism and fishing in a stage-structured population model of the blue crab in Chesapeake Bay}
\date{}

\author[1,2]{Fangming Xu}
\affil[1]{Department of Mathematics, William \& Mary, Williamsburg, Virginia 23187-8795}
\affil[2]{Department of Statistics and Data Science, Yale University, New Haven, Connecticut 06511\\
fangming.xu@yale.edu}

\author[1,3]{Leah B.~Shaw}
\affil[3]{lbshaw@wm.edu}

\author[1,4]{Junping Shi}
\affil[4]{jxshix@wm.edu}

\author[5]{Romuald N. Lipcius}
\affil[5]{Virginia Institute of Marine Science, William \& Mary, P.O. Box 1346, Gloucester Point, Virginia 23062\\rom@vims.edu}

\maketitle

\begin{spacing}{1.9}
\newpage
\begin{abstract}

The blue crab (\textit{Callinectes sapidus}) is a dominant ecological species of high commercial value along the western Atlantic Ocean and Gulf of Mexico. Spawning stock and recruitment of the Chesapeake Bay population declined by 80\% in the 1990s, while average annual female abundance dropped 50\% from 172 million crabs in 1989-1993 to 86 million crabs in 1994-2007. After severe management actions were implemented in 2008, annual female abundance rebounded to pre-1994 levels, and stabilized at an average of 161 million crabs in 2008-2019. The stepwise decline of the population in the early 1990s, followed by a consistently low level of abundance for 15 y and a jump to high abundance after 2008, suggested the existence of alternative stable states or a regime shift. Alternatively, high fishing pressure combined with unexpectedly low recruitment in 1992 could have triggered a proportional decline in the population, followed by a population increase in 2008 proportional to rigorous management actions that reduced fishing pressure. We evaluated these alternatives with a stage-structured dynamic population model using ordinary differential equations. In addition, stock assessment models for the blue crab assume that exploitation rates due to fishing and mortality rates due to predation and cannibalism are independent of density. Hence, we also investigated the role of density-dependent predation, cannibalism and fishing in blue crab population dynamics. Based on our simulations, we conclude that for the blue crab population in Chesapeake Bay: (1) bistable positive states are not likely with biologically realistic parameter values; (2) hyperbolic (depensatory) fishing will not produce extinction at the range of juvenile and adult densities observed in the bay; and (3) crabs can survive a higher fishing rate under the more realistic assumption of sigmoidal (density-dependent) predation and cannibalism than under constant (density-independent) predation and cannibalism. The blue crab population in Chesapeake Bay is therefore more resistant to fishing due to density-dependent predation and cannibalism than would be expected from density-independent predation and cannibalism; depensatory fishing is unlikely to lead to fishery collapse; and changes in fishing pressure will probably not produce disproportional changes in population abundance. These collectively indicate that the blue crab population in Chesapeake Bay is resilient to a range of biotic and abiotic disturbances.

\end{abstract}

\noindent\textbf{Keywords}: blue crab restoration, differential equation model, stage-structured, alternative stable states, threshold

\newpage

\section*{Introduction}

The blue crab (\textit{Callinectes sapidus}) is a widespread species in marine and estuarine habitats along coasts of the western Atlantic Ocean and Gulf of Mexico \citep{Churchill1919}, where it also supports valuable commercial and recreational fisheries \citep{Kennedy2007}. For instance, in 2019 annual commercial landings of blue crab were 66,497 T valued at US\$205.6 million \citep{NOAA2019}. The Chesapeake Bay supports one of the largest populations and fisheries of blue crab in the United States, constituting 38\% of the total catch in the Atlantic and Gulf coasts \citep{NOAA2019}. In addition, the blue crab is a dominant species ecologically, serving as both predator and prey in the Chesapeake Bay food web \citep{Hines2007, Lipcius2007}.

Early in the 1990s the blue crab spawning stock and recruitment in Chesapeake Bay declined by 80\% \citep{Lipcius2002}, while average annual female abundance dropped 50\% from 172 million crabs in 1989-1993 to 86 million crabs in 1994-2007 \citep{MDNR2019}. The sharp decline resulted in a range of management and recovery actions through 2007 \citep{Miller2011}, including establishment of an extensive spawning sanctuary covering about 75\% of the spawning grounds \citep{Lipcius2001sanctuary,Seitz2001sanctuary,Lipcius2003}, which proved effective in reducing exploitation of the spawning stock \citep{Lambert2006}. The sanctuary and other management actions likely prevented collapse of the stock \citep{Lambert2006} but were insufficient to stimulate recovery \citep{Miller2011}. Consequently, severe fishery reductions were implemented in 2008 by management agencies, including the Virginia Marine Resources Commission, Potomac River Fisheries Commission, and Maryland Department of Natural Resources (MDDNR), leading to a 34\% reduction in female landings across Maryland and Virginia \citep{Miller2011} and triggering population recovery. Since 2008, annual female abundance rebounded to pre-1994 levels, and stabilized at an average of 161 million crabs in 2008-2019 \citep{MDNR2019}.

The stepwise decline of the spawning stock and recruitment in the early 1990s \citep{Lipcius2002}, followed by a consistently low level of abundance for 15 y and a jump to high abundance after 2008 \citep{Miller2011}, suggested the existence of alternative stable states or a regime shift. Alternatively, high fishing pressure combined with unexpectedly low recruitment in 1992 could have triggered a proportional decline in the population \citep{Lipcius2002}, followed by a population increase in 2008 proportional to rigorous management actions that reduced fishing pressure on the spawning stock \citep{Miller2011}. These alternatives are represented by two main hypotheses for population changes since 1990. 

The first hypothesis is the existence of bistable positive states. If the blue crab population exhibits two positive, stable equilibrium states for a given fishing rate, the population decline in the early 1990s could be explained as a plunge from the high stable state to the low stable state. Similarly, the sudden jump in the population in 2008 could be explained as a shift from the low stable state to the high stable state. The second hypothesis is that there is only one positive equilibrium for a given fishing rate, and that population abundance is inversely proportional to fishing pressure. Complementary to the second hypothesis is the supposition that management actions from the early 1990s through 2007, such as implementation of the spawning sanctuary, may have saved the blue crab from local extinction and maintained the population at the low stable state.

In addition, current stock assessment models for the blue crab assume that exploitation rates due to fishing and mortality rates due to predation and cannibalism are independent of density \citep{Miller2011,MDNR2019}, despite the prevalence of density-dependent phenomena in marine populations \citep{Lorenzen2019}. Hence, we also investigated the role of density-dependent predation, cannibalism and fishing in blue crab population dynamics. To evaluate the hypotheses and effects of density dependence in the blue crab population of Chesapeake Bay, we developed a dynamic population model using a system of ordinary differential equations.

\section*{Methods} \label{model}

\subsection*{Life Cycle} \label{life cycle}

\begin{figure}[ht!]
 \begin{center}
\includegraphics[height=1.85in,width=6.0in,angle=0]{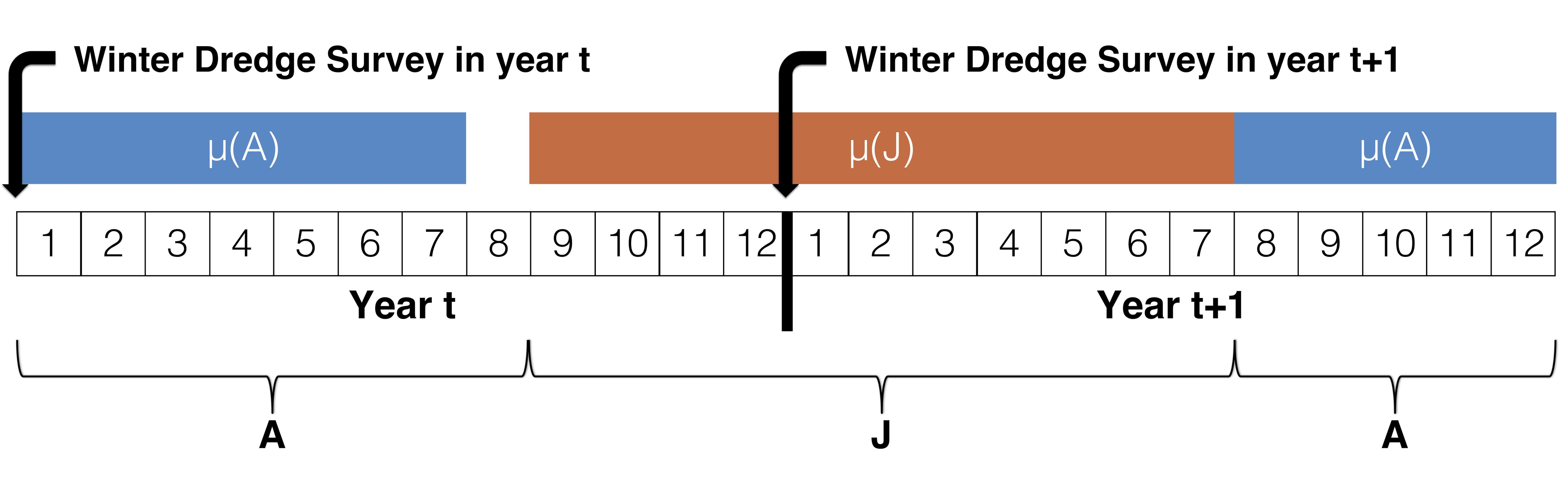}
 \caption{Life cycle of blue crab in a two-year period. $A$: adult stage; $J$: juvenile stage; $\mu (A)$: adult mortality; $\mu (J)$: juvenile mortality.}
 \label{fig:LifeCycle}
\end{center}
\end{figure}

The demographic model for blue crab population dynamics consists of two ordinary differential equations (ODEs), one for adults $A$ and a second for juveniles $J$. The set of ODEs accounts for reproduction, predation, cannibalism, maturation, natural mortality, and fishing mortality. In Figure~\ref{fig:LifeCycle} the life cycle of the blue crab is depicted during a two-year period. We assume that egg-bearing females, which have carried an egg mass for about 2 weeks, hatch larvae on average during August 1 of year $t$ \citep{Lipcius2002,Smith2007}. Larvae are advected outside of Chesapeake Bay, where they remain in the plankton for about 1 month before reinvading the bay \citep{vanMontfrans1995,Epifanio2007}. Once in the bay, the postlarval stage settles in nursery habitats and metamorphoses to the first juvenile instar \citep{Metcalf1992,Lipcius2007}. For simplicity, we do not include larval stages in our model. Juveniles suffer 11 months of mortality ($\mu (J)$ in Figure~\ref{fig:LifeCycle}) from predation by fish and cannibalism by adult crabs \citep{Hines2007,Lipcius2007}. Starting in August of the next year, blue crabs have reached the adult stage and now experience natural mortality and fishing mortality. We denote the sum of natural mortality and fishing mortality as $\mu (A)$ (Figure~\ref{fig:LifeCycle}). The life cycle diagram also indicates how $J$ and $A$ are classified in the model, which is based on whether crabs are able to reproduce. The transition from $J$ to $A$ was placed at August 1, which is about the midpoint of the spawning season \citep{Lipcius2003}.

\subsection*{Model Equations} \label{equation}

Our model is based on adult female density and juvenile female density. As described above, crabs are considered juveniles from September in the first year through July in the second year, before the midpoint of the spawning season. They are considered adults 11 months after their entry into the juvenile stage. Adults $A(t)$ and juveniles $J(t)$ in year $t$ are measured in numbers of female crabs per 1000 m$^2$, following the standard for the Blue Crab Winter Dredge Survey \citep{Sharov2003}.

The rate of change in density of juveniles is represented by an ordinary differential equation:
\begin{equation}\label{1}
\frac{dJ}{dt} = R(A) - P(A,J) -\beta J,
\end{equation} 
where the first term $R(A)$ represents recruitment of juvenile crabs into the population, the second term $P(A,J)$ represents predation and cannibalism, and the third term $\beta J$ represents maturation of juveniles.

\begin{figure}[ht!]
    \begin{center}
    \includegraphics[scale=0.38]{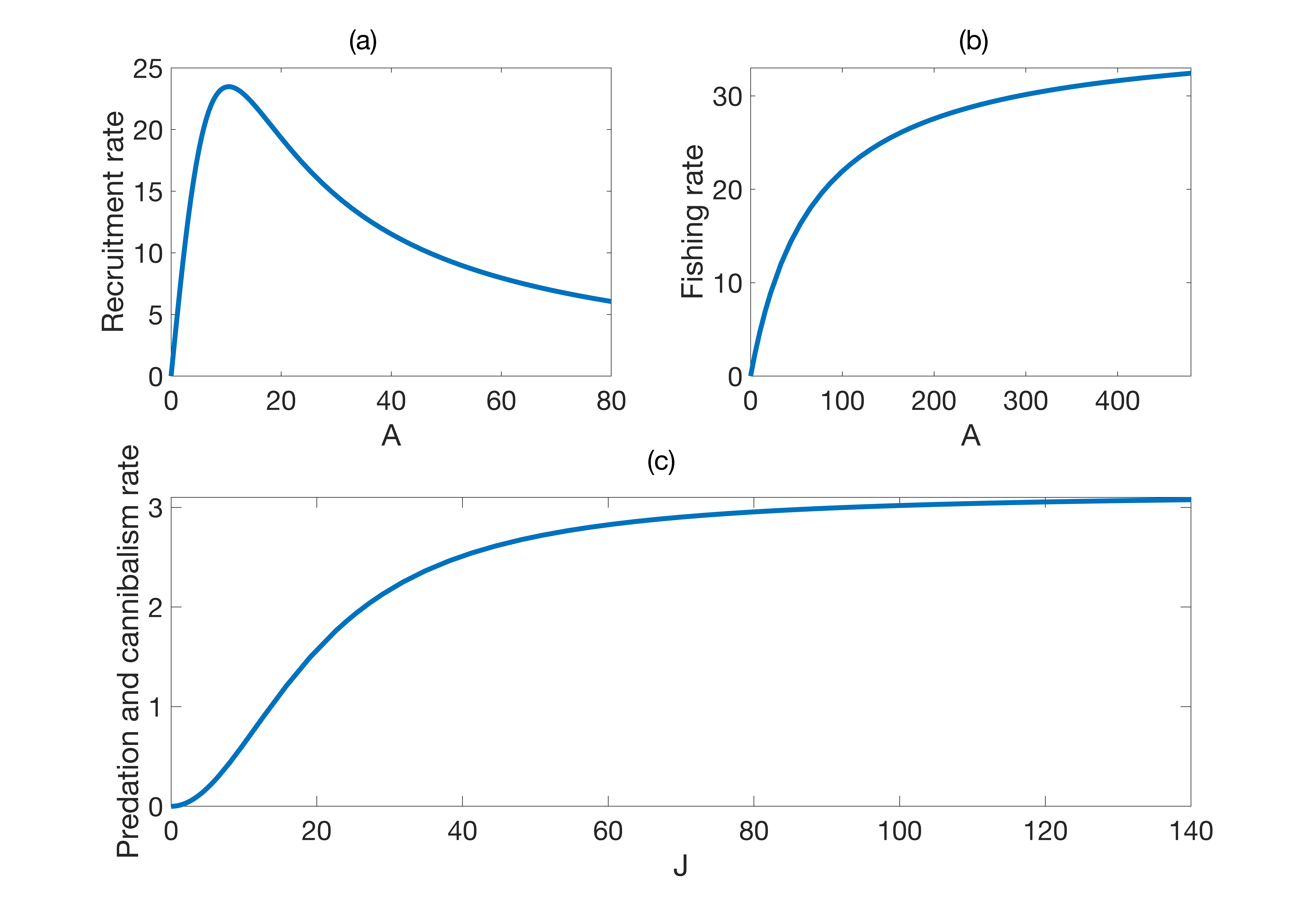}
    \caption{Functions used in the model: (a) Ricker stock-recruitment function; (b) Hyperbolic fishing function; (c) Sigmoidal predation and cannibalism function.}
    \label{fig:Functions}
    \end{center}
\end{figure}

The recruitment function $R(A)$ can be of several different forms. Table \ref{tab:Recruitment} shows the 6 candidate models. We choose the best model based on data fitting below. Of all 6 models, a Ricker model (\texttt{rickerm}, $R(A)=\ds\frac{\alpha A}{1+bA^2}$) is more likely to produce bistability than other models. The Ricker model has a unimodal shape where $\alpha$ is the maximum per capita rate for adults to produce juveniles and $b$ controls the decrease in reproduction at high adult densities (Figure \ref{fig:Functions}(a)). As the density of adults approaches infinity, $1+bA^2$ goes to infinity, and the recruitment term goes to 0. This implies that when there are too many adults, competition for food and resources intensifies, and reproduction rate decreases.

Predation and cannibalism of juvenile blue crabs is represented by $P(A,J)$, which can be either (1) linear in $J$, $P(A,J)=lJ$, or (2) sigmoidal in $J$, $P(A,J)=(p+A) \ds\frac{k_{max}J^2}{x^2+J^2}$. For the linear form, $l$ represents a linear mortality for juveniles due to predation and cannibalism. The sigmoidal term represents a type III functional response as in \citet{Moksnes1997}, who determined that the cannibalism function is of the form $A \ds \frac{k_{max}J^2}{x^2+J^2}$, where $k_{max}$ is the maximum feeding rate and $x$ is the density of prey at which the feeding rate is half of the maximum (Figure \ref{fig:Functions}(c)). When juvenile density is low, the cannibalism rate is also low because the encounter rate with juveniles is minimal, such that adult crabs switch to more abundant prey such as clams and oysters. When juvenile density increases, the cannibalism rate also increases, but will eventually approach an asymptote due to maximal handling times or satiation. Specifically, when juvenile crabs are abundant, cannibalizing adult crabs can locate juveniles easily and eat as many juveniles as they can until they reach satiation. Cannibalizing crabs are also limited by handling time, which includes capture, consumption and digestion of juveniles, such that their feeding rate is bounded. Since generalist predators exhibit a similar functional response as cannibalizing crabs \citep{Hines2007}, we represent predation by simply multiplying $ \ds\frac{k_{max}J^2}{x^2+J^2}$ by a parameter $p$ for the density of predators. Thus, the combined predation and cannibalism terms are $(p+A) \ds\frac{k_{max}J^2}{x^2+J^2}$. 

The maturation term $-\beta J$ represents maturation of juveniles. As juveniles grow larger, they leave the stage $J$ at a maturation rate $\beta$, where $\beta = \ds\frac{1}{\text{maturation  \  time}}$.  

The rate of change in the density of adult females is represented by another ordinary differential equation:
\begin{equation}\label{2}
\frac{dA}{dt} = \beta J-mA - F(A).
\end{equation}
where the rate of change in adult female density is composed of the crabs that join the adult class, natural mortality of adults and a fishing mortality $F(A)$.

The new adult term is derived directly from the juvenile equation. As crabs grow larger and mature, they leave the juvenile stage and add to the adult stage. We assume that all juvenile crabs that survive predation and cannibalism join the adult stage after 11 months. In the natural mortality term $-mA$, $m$ is the natural mortality rate, which includes mortality of adult females caused by disease, old age, and any other natural sources of mortality. 
    
We also include a fishing mortality function $-F(A)$ in the adult equation. Here, $F(A)$ can be of two different forms: (1) linear, $F(A)=f_l A$, or (2) hyperbolic, $F(A)=\ds\frac{f_h A}{s+A}$. The parameter $f_l$ represents linear fishing mortality, $f_h$ represents maximum fishing mortality for hyperbolic fishing, and $s$ is the adult density at half of the maximum fishing rate (Figure \ref{fig:Functions}(b)). Linking Equations \eqref{1} and \eqref{2}, the population dynamics of juvenile and adult crabs is described by a system of two ordinary differential equations:
\begin{equation}\label{3}
\begin{split}
\ds\frac{dJ}{dt} &= R(A) 
-(p+A) \frac{k_{max}J^2}{x^2+J^2}-\beta J,\\
\frac{dA}{dt} &= \beta J-mA - F(A).
\end{split}
\end{equation}
with parameters and their units detailed in Table \ref{tab:Parameters}.

\subsection*{Estimation of parameters} \label{estimation}

\subsubsection*{Abundance}

\begin{figure}[ht!]
    \begin{center}
    \includegraphics[scale=0.48]{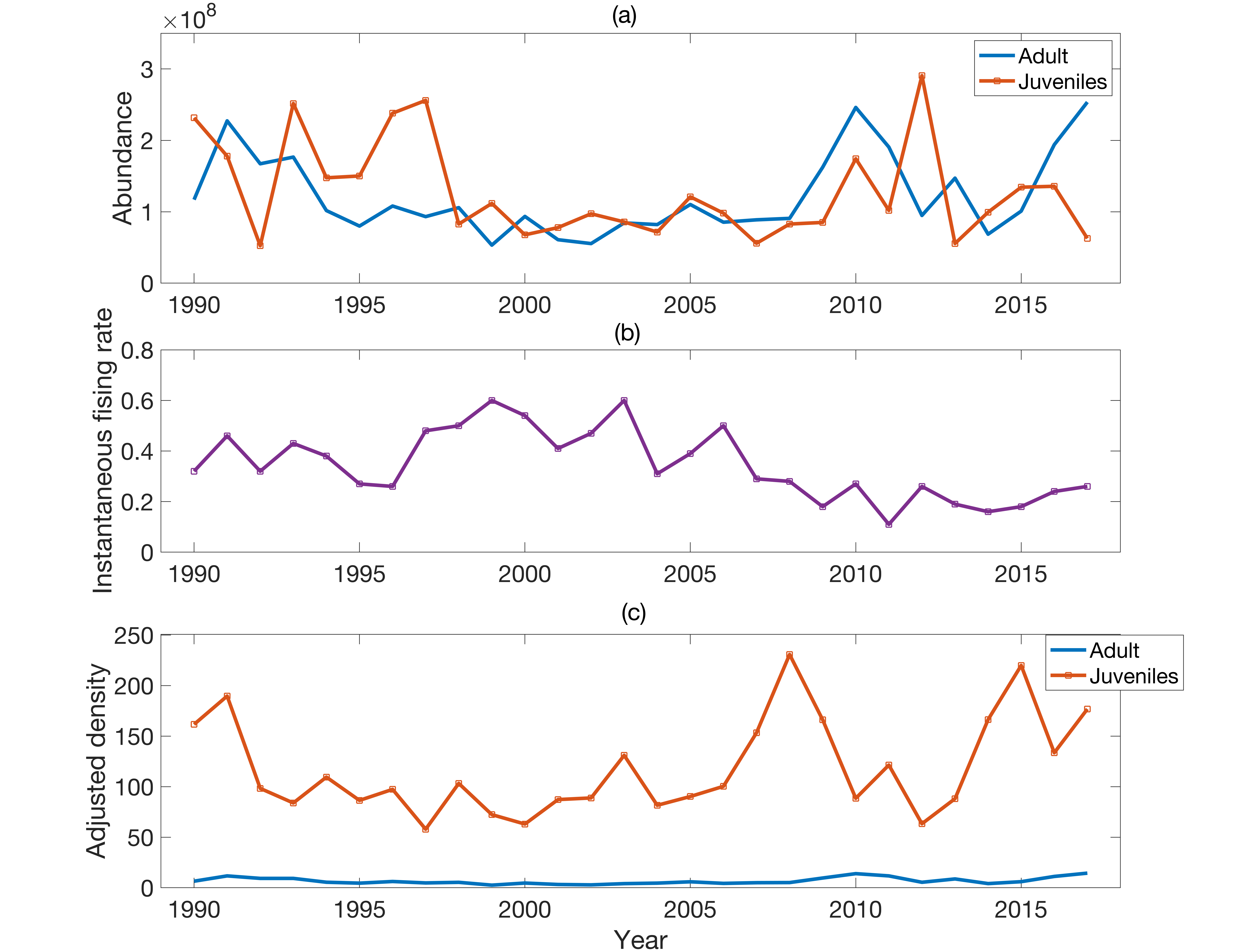}
    \caption{(a): Total abundance of adult and juvenile crabs in the Chesapeake Bay (females only); (b): Instantaneous fishing rate for adults; (c): Adjusted density of spawning adults on August 1 and corresponding juveniles on September 1.}
    \label{fig:TimeSeries}
    \end{center}
\end{figure}

The total number of adult female crabs in Chesapeake Bay was estimated from Winter Dredge Survey (WDS) data and the annual estimates of exploitation rate from 1990 to 2017 \citep{MDNR2019}. The raw data of the number of adult and juvenile female crabs is displayed in Figure \ref{fig:TimeSeries}(a). The total area of the Bay sampled by the survey is 9,812 km$^2$, so we converted the WDS data to number per 1000 m$^2$ to be consistent with density units used by the WDS \citep{Sharov2003,MDNR2019}.

Since adults on August 1 of year $t$ produce the juveniles on September 1 of year $t$ (Figure \ref{fig:LifeCycle}), we need to calculate their density to parameterize the recruitment term in our model. We estimate adult female density by calculating forward from the WDS data on January 1 in year $t$ to August 1 in year $t$. We calculate juvenile density on September 1 in year $t$ back from the January 1 year $t+1$ WDS juveniles. Density of juvenile female crabs (Figure \ref{fig:TimeSeries}(a)) must be modified because juvenile density is underestimated by the WDS \citep{Ralph2013}. From the life cycle diagram (Figure \ref{fig:LifeCycle}), we note that adults measured by the WDS on January 1 of year $t$ will suffer from 7 months of adult natural mortality $m= 0.9$ $y^{-1}$ \citep{Hewitt2007} and fishing mortality before they become spawning females in August. Yearly instantaneous fishing rates (Figure \ref{fig:TimeSeries}(b)) are calculated from the harvest data in the WDS \citep{MDNR2019}.

\begin{figure}[ht!]
    \begin{center}
    \includegraphics[scale=0.45]{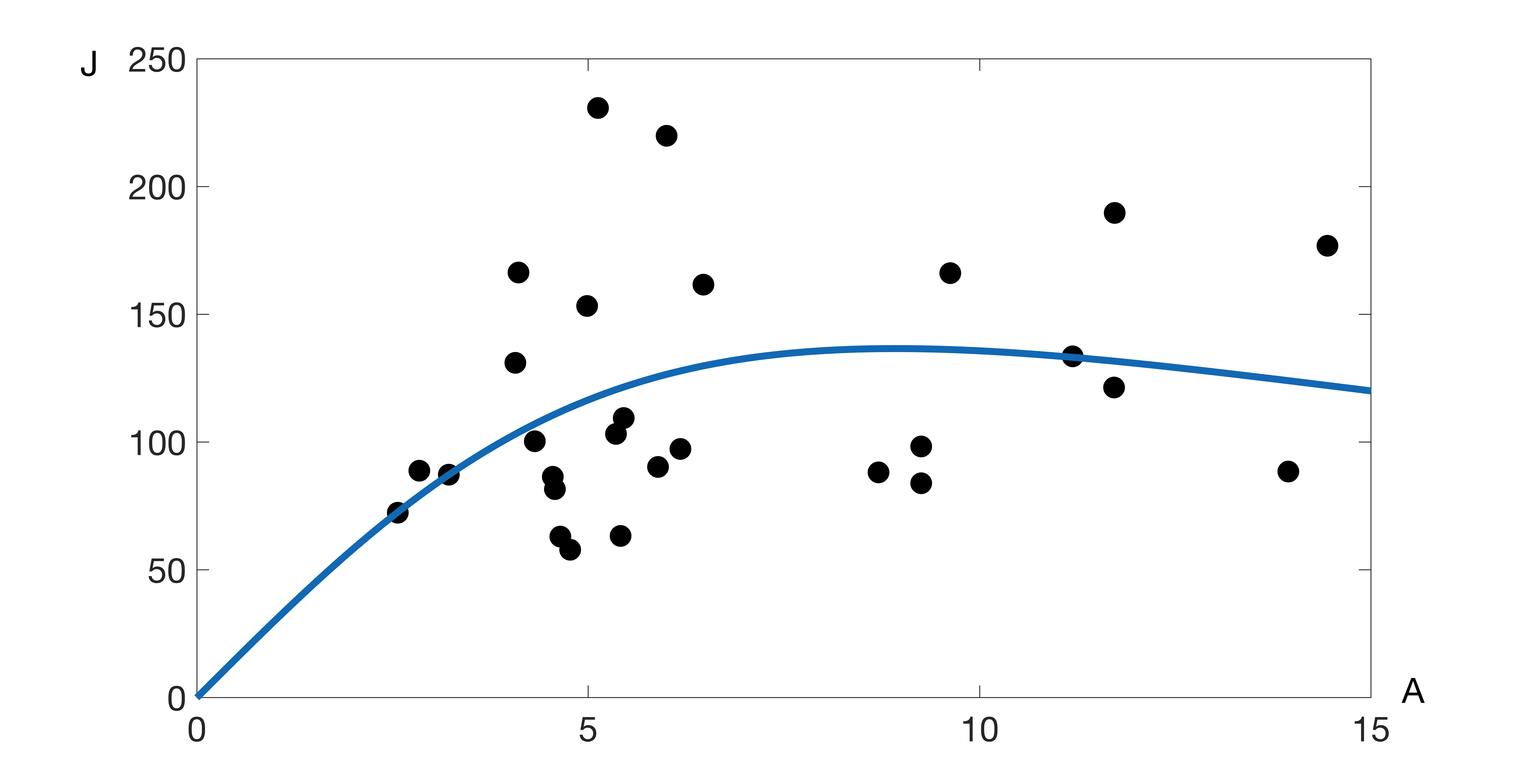}
    \caption{Recruitment fitted from adjusted spawning adults and juveniles using Matlab ''cftool'' toolbox.}
    \label{fig:Recruitment}
    \end{center}
\end{figure}

Annual fishing mortality rates can be as high as 0.6 $y^{-1}$ (Figure \ref{fig:TimeSeries}(b)). Together with natural mortality of 0.9 $y^{-1}$ \citep{Hewitt2007}, we have an estimate of total mortality of adults per year. Adults measured at the beginning of January in year $t$ will suffer from 7 months of total adult mortality before they become the spawning stock on August 1, so we discount the WDS data at year $t$ by 7 months of adult total mortality. Next, spawning females hatch larvae which develop and become juvenile crabs on September 1. Juveniles suffer from 4 months of juvenile mortality of 1.5 $y^{-1}$ until January 1 in year $t+1$ , when the WDS in year $t+1$ takes place. [The rate of 1.5 $y^{-1}$ is based on the relationship between juvenile survival and size \citep{Bromilow2017} relative to an adult mortality rate of 0.9 $y^{-1}$ \citep{Hewitt2007}.] Thus, only a portion of juveniles will survive and be surveyed at year $t+1$. We increased the $F0_{t+1}$ (juvenile females surveyed in year $t+1$) accordingly to estimate juveniles that entered the juvenile stage on August 1 of year $t$. Note that the larval stage is not parameterized in this model. Adjusted juvenile female densities are shown in Figure \ref{fig:TimeSeries}(c). We then fitted the Ricker recruitment term with these data using the \texttt{cftool} toolbox in \texttt{Matlab}, resulting in best fitting parameters $\alpha=30.62$ and $b=0.01256$. The data and fitted curve are shown in Figure \ref{fig:Recruitment}.

Next, we assume that juvenile crabs mature in 11 months. They are considered juveniles from August 1 in year $t$ to July 31 year $t+1$ (Figure \ref{fig:LifeCycle}). Thus the maturation rate is $\beta=\ds\frac{1}{\text{maturation \ time}} = 1.09$ $y^{-1}$. 

\subsubsection*{Fishing mortality function}

\begin{figure}[ht!]
   \begin{center}
   \includegraphics[width=1\linewidth]{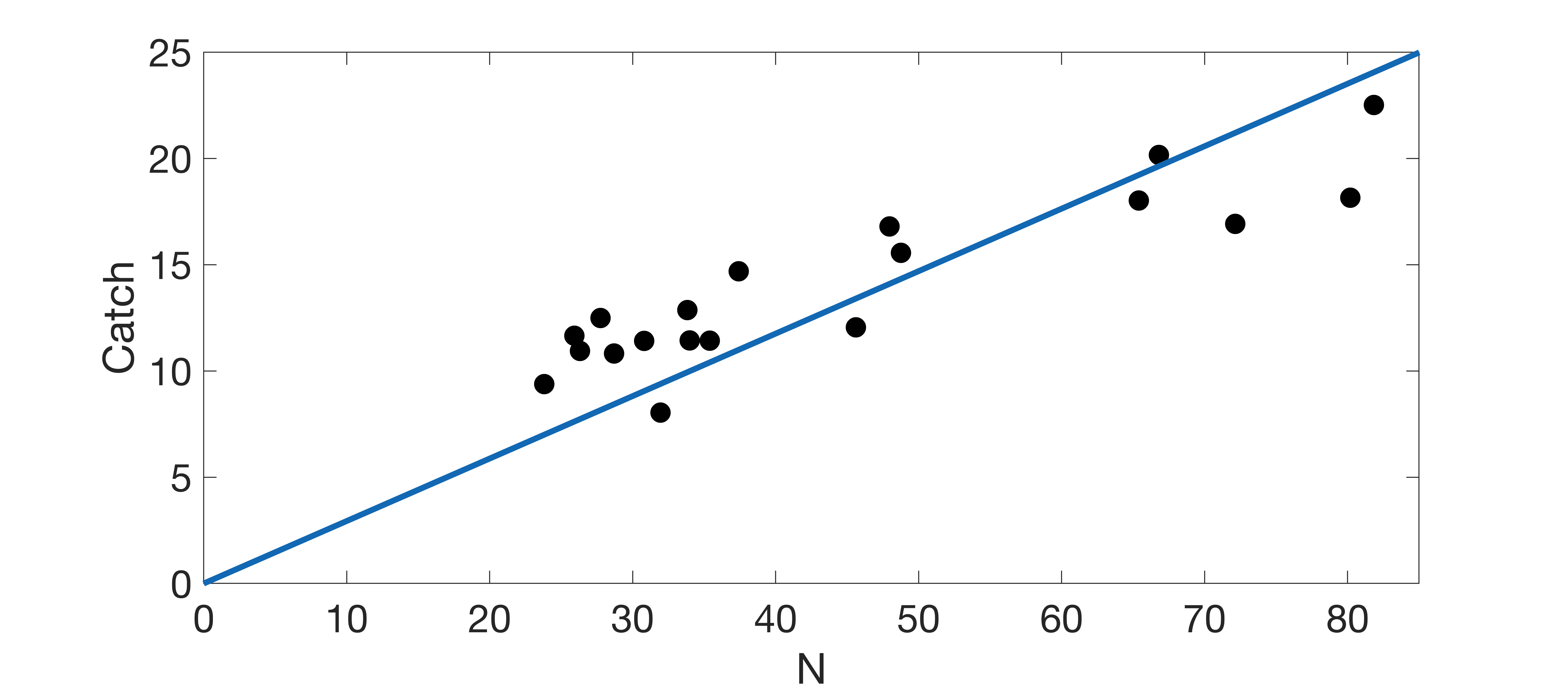}  
  \caption{Data and fitted curve for linear fishing term.}
   \label{fig:Linear}
   \end{center}
\end{figure}

\begin{figure}[ht!]
  \begin{center}
  \includegraphics[height=3.07in,width=4.0in,angle=0]{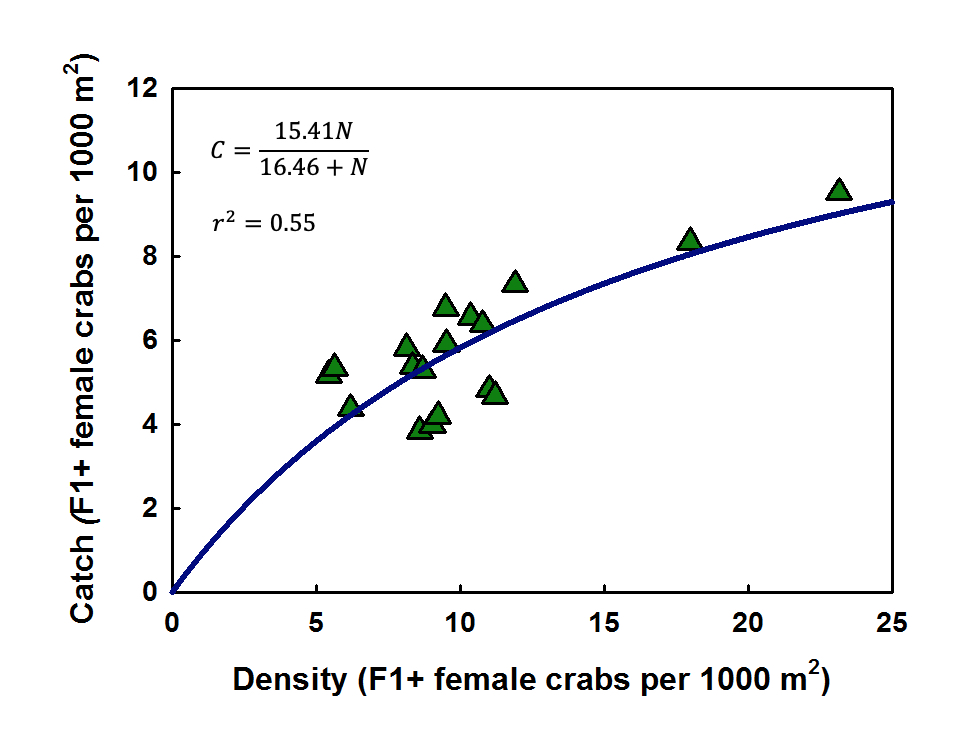}  
  \caption{Data and fitted hyperbolic curve for fishing term. The curve was fit using nonlinear regression with the Levenberg-Marquardt algorithm in SigmaPlot 14.}
 \label{fig:Hyperbolic}
\end{center}
\end{figure}

For the fishing term $-F(A)$, there are two possible forms: linear and hyperbolic. From WDS data, total female abundance ($N$) and number of crabs caught each year ($Catch$) are analyzed from 1990 to 2008 \citep{MDNR2019}. The units for total abundance and catch are in million crabs baywide, which we converted to number of crabs per thousand square meters to be consistent with our model. Linear density-independent fishing is in the form $F(A)=f_l A$. By fitting the data for a line starting from the origin, we estimate $f_l$ at 0.294. The data and fitted curve are shown in Figure \ref{fig:Linear}. Hyperbolic fishing has the form $F(A)=\ds\frac{f_h A}{s+A}$. The best fit for the hyperbolic model was $f_h=37.097$ and $s=69.130$, estimated using the \texttt{cftool} toolbox in \texttt{Matlab}. The data and fitted curve for hyperbolic fishing are shown in Figure \ref{fig:Hyperbolic}.

\subsubsection*{Predation and cannibalism}

In the term for predation and cannibalism of juveniles in \eqref{1}, we estimated parameter values of $p$, $k_{max}$ and $x$. Two lines of evidence indicate that cannibalism follows a sigmoid functional response (i.e., number of prey killed per predator as a function of prey density). First, the blue crab is a generalist predator that switches from sparse prey to abundant prey \citep{Lipcius1986,Hines2007,Lipcius2007}. Second, \citet{Moksnes1997} demonstrated a sigmoid functional response based on large blue crabs cannibalizing juvenile crabs in artificial seagrass in lab experiments. Furthermore, given that the main predators of juvenile blue crabs are generalist predators \citep{Hines2007,Lipcius2007}, we assume that the functional response of predators to juvenile blue crab prey is also sigmoid.

We parameterized the sigmoidal predation and cannibalism function so that at typical adult and juvenile densities, the per capita mortality of juveniles is 1.5 $y^{-1}$, as noted above. Setting $P(J,A)=1.5J$, and with Ricker recruitment, linear predation and cannibalism, and linear fishing, the positive equilibrium crab densities in \eqref{3} are $J_{lin}=30.58$ and $A_{lin}=27.92$. These are within the range of the adjusted adult female densities (Figure \ref{fig:TimeSeries}(c)), supporting the assumption that 1.5 $y^{-1}$ is a realistic mortality rate for juveniles. We then used this linear mortality rate to estimate sigmoidal predation and cannibalism by setting $\ds (p+A_{lin})\frac{k_{max}J_{lin}}{x^2+J_{lin}^2}=1.5$ to obtain $k_{max}=\ds \frac{1.5 (x^2+J_{lin}^2)}{J (p+A_{lin})}$. We also assumed that the predation rate will be similar to the cannibalism rate \citep{Hines2007}, and we approximated $p$ using the steady state $A_{lin}=27.92$. Hence we estimated $k_{max}=\ds\frac{3 (x^2+J_{lin}^2)}{4 J_{lin} A_{lin}}$.

The parameter $x$ is the juvenile density at which prey switching occurs (i.e., at which predation on juveniles is half of its maximum value). Because prey switching is observed \citep{Hines2007}, we expect $x$ to be in the range of the observed juvenile densities. According to the life cycle (Figure \ref{fig:LifeCycle}), the highest juvenile density will occur on September 1 in year $t$, when juveniles are first recruited, and which we estimated previously (Figure \ref{fig:TimeSeries}(c)). Average juvenile density on September 1 is about 118 crabs per 1000 m$^2$ , which gives an upper bound for $x$. The lowest juvenile density occurs on July 31 in year $t+1$, immediately before juveniles become adults. Juveniles experience 11 months of juvenile mortality at an average rate of 1.5 $y^{-1}$ from their maximum on September 1 of year $t$ to their minimum on July 31 of year $t+1$. We discount juvenile density on September 1 by a factor of $\exp(- 1.5 \cdot 11/12)$, which yields an average lowest juvenile density of about 30 crabs per 1000 m$^2$. Thus, 30 to 118 is a range for $x$. Substituting all the estimates $31\leq x \leq 118$, $J_{lin}=30.58$ and $A_{lin}=p=27.92$ into the function $k_{max}=\ds\frac{3 (x^2+J_{lin}^2)}{4 J_{lin} A_{lin}}$, we obtain a range of $k_{max}$ between 1.61 and 13.05.

\subsubsection*{Spawning stock-recruitment function}

Using our estimates for adult and juvenile densities, we fit \texttt{linear}, \texttt{hyperbolic} and \texttt{Ricker} functions to the data using the \texttt{nls} function (nonlinear least squares) in \texttt{R} \citep{R2017}. A \texttt{null} model with constant recruitment was used as a comparison to other functions. Also, when adult crab density is low, finding a mate may be difficult, so the per capita recruitment rate might decrease at low population densities (i.e., Allee effect). Hence, each of the nonlinear models was analyzed with and without an Allee effect. We also assumed that recruitment = 0 when abundance of adults = 0. Thus, all of the models except the null model have a y-intercept = 0 (Table \ref{tab:Recruitment}).

To determine the best-fitting model, the 6 statistical models (Table \ref{tab:AICc}) representing alternative functions and hypotheses \citep{Chamberlin1890} were evaluated following an Information Theoretic approach \citep{Burnham2002, Anderson2008}. The suite of models included the sparse (\texttt{null}) model. Akaike Information Criterion (AIC) values from each model were used to calculate AIC$_c$, a second-order bias correction estimator \citep{Anderson2008}. Model probabilities ($w_i$), based on $\Delta_i$ values, were used to rank the different models against the model with the lowest AIC$_c$, and estimate the probability that a particular model was the best model. Any model with $w_i$ less than 0.10 was eliminated from further consideration \citep{Anderson2008}. The coefficient of discrimination ($r^2$) and likelihood ratio $\chi^2$ tests \citep{Vuong1989} were used to assess model fit.

\begin{figure}[ht!]
    \begin{center}
    \includegraphics[scale=0.8]{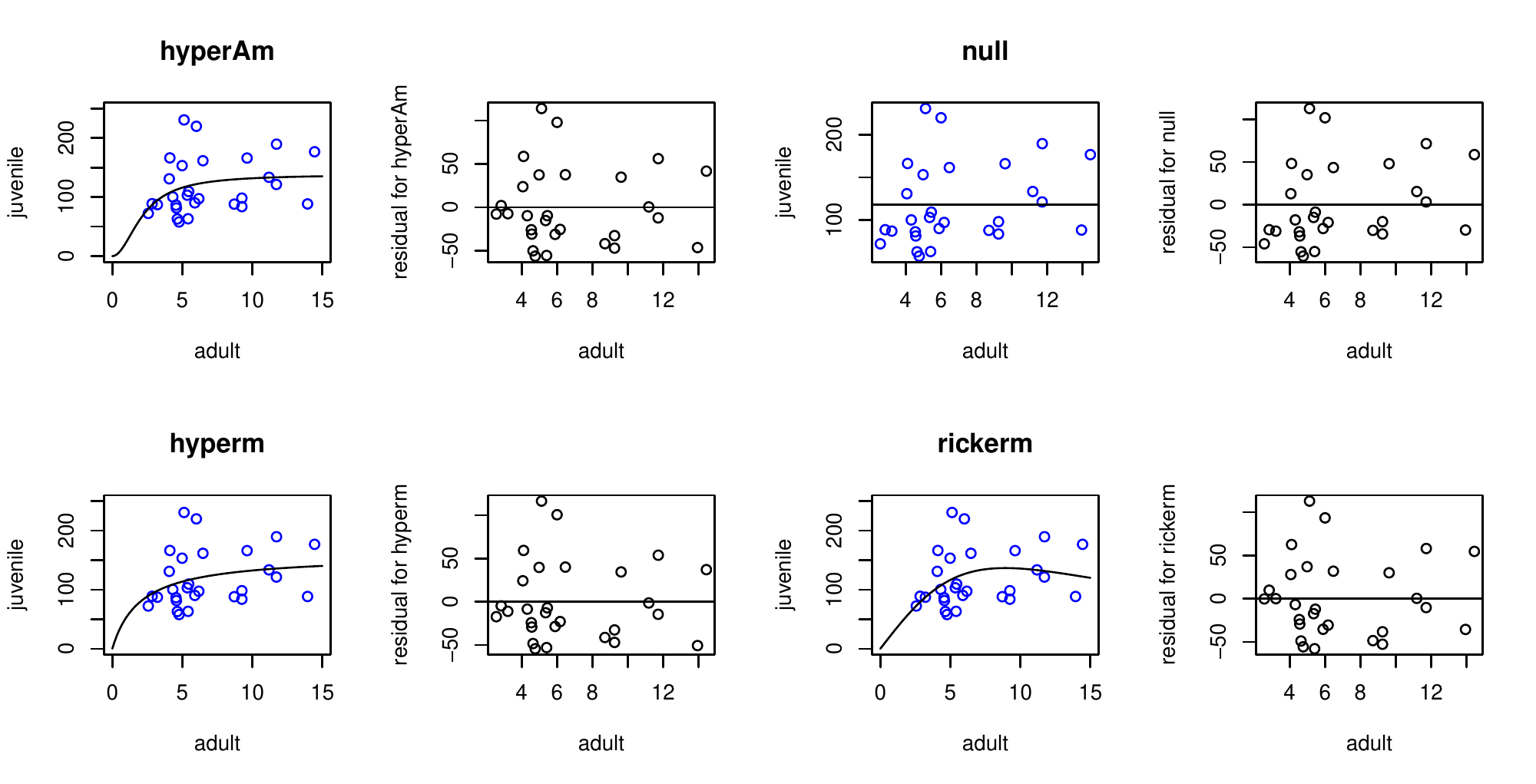}
    \caption{Four candidate models and their residual plots. The units for the axes are density per 1000 m$^2$. Blue dots are the actual yearly data for the spawning stock and the recruits. Black curves represent the predicted recruits based on the fitted models. For the residual plots, black dots represent the residual points.}
    \label{fig:Resids}
    \end{center}
\end{figure}

The \texttt{hyperAm} (hyperbolic with Allee effect) model produced the best fit with $w_i$ = 0.268 (Table \ref{tab:AICc}). The null, \texttt{hyperm} (hyperbolic without an Allee effect) and \texttt{rickerm} (Ricker without an Allee effect) models had $w_i >$ 0.10 and also merited further consideration. The other two models were excluded from consideration because their weighted probabilities ($w_i$) were $<$0.01. In the analysis of residuals, those of \texttt{hyperAm} and \texttt{rickerm} appeared random, whereas those of the \texttt{null} and \texttt{hyperm} models were non-random at low adult densities (Figure \ref{fig:Resids}), so we excluded the latter. Based on the AIC$_c$ and residual plots, we concluded that \texttt{hyperAm} was the best model for the recruitment term, though \texttt{rickerm} also performed well. Since the Ricker function is the most likely form to produce bistability, we conducted the mathematical analysis with \texttt{rickerm} as well as with \texttt{hyperAm}.

\section*{Results and Discussion} \label{results}

To understand the population dynamics of blue crabs under different conditions, we investigated three different forms of the model: the simplest one with constant predation and cannibalism rates and linear fishing mortality, the slightly more complex one with density-dependent predation and cannibalism rates and linear fishing mortality, and the most complex one with density-dependent predation and cannibalism rates and hyperbolic fishing mortality.

\subsection*{Basic Dynamics}

Here we describe basic dynamics of the model \eqref{3} under general assumptions for the recruitment function $R(A)$ and fishing mortality function $F(A)$:

\begin{enumerate}
    \item[(R)] The function $R(A)$ satisfies $R(0)=0$, $R(A)>0$ for $A>0$, and $R_M=\ds\max_{A\geq 0}R(A)<\infty$.
    \item[(F)] The function $F(A)$ satisfies $F(0)=0$, $F(A)\geq 0$ and $F'(A)\geq 0$ for $A>0$.
\end{enumerate}
Note that for the six functions listed in Table \ref{tab:Recruitment}, all except \texttt{null} satisfy (R), and both the linear and hyperbolic fishing functions satisfy (F).

Under the assumptions (R) and (F), a solution $(J(t),R(t))$ of \eqref{3} with non-negative initial value is bounded: $0\leq J(t)\leq R_M/\beta$ and $0\leq A(t)\leq R_M/m$, and any equilibria of \eqref{3} are also in that range. Also from the Bendixson Criterion \citep{steven1994nonlinear}, there is no limit cycle for \eqref{3}, and from Poincar\'e-Bendixson theory \citep{steven1994nonlinear}, any solution converges to a non-negative equilibrium asymptotically when $t\to\infty$.  

The extinction equilibrium $(J,A)=(0,0)$  is locally asymptotically stable if $m+F'(0)>R'(0)$; i.e., the combined natural and fishing moralities are larger than recruitment when the population size is small. For the recruitment functions with Allee effect (\texttt{hyperAm}, \texttt{rickerAm}), this is naturally true as $R'(0)=0$. For other recruitment functions without Allee effect (\texttt{linearm}, \texttt{hyperm}, \texttt{rickerm}), $(0,0)$ is  locally asymptotically stable if $m+f>\alpha$, where $f=f_l$ or $f_h$ depending on whether a linear or hyperbolic fishing function is used. If $mA+F(A)>R(A)$ holds for all $A\geq 0$, then $(0,0)$ is the only equilibrium so the population will become extinct. 

Model \eqref{3} may have one or multiple positive equilibria. Assuming equilibrium, solving $J$ from the second equation of \eqref{3}, we have $J=(mA+F(A))/\beta$; and solving $J$ from the sum of the two equations of \eqref{3}, we have $J=x\sqrt{G(A)/(1-G(A))}$, where $G(A)=\ds\frac{R(A)-mA-F(A)}{k_{max}(p+A)}$. Hence for any positive equilibrium $(J,A)$, the adult density $A$ satisfies 
\begin{equation}\label{4}
    \frac{mA+F(A)}{\beta}=x\sqrt{\frac{R(A)-mA-F(A)}{k_{max}(p+A)-R(A)+mA+F(A)}}.
\end{equation}

\subsection*{Bistability}

\begin{figure}[ht!]
\begin{subfigure}{0.9\textwidth}
  \centering
  \includegraphics[width=\textwidth]{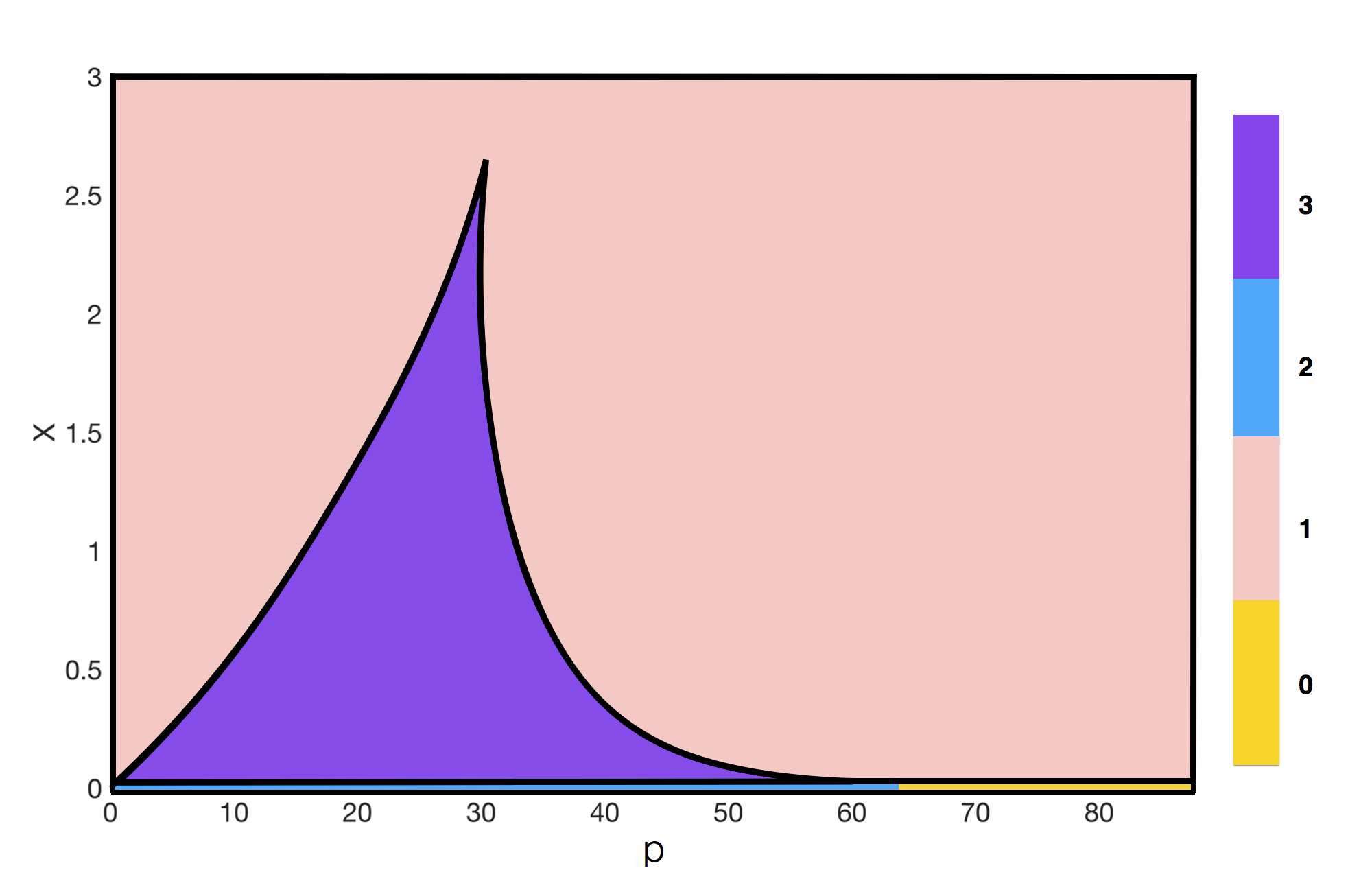}
 \caption{Number of positive equilibria  with different $p$ and $x$ values.}
\end{subfigure}
\begin{subfigure}{0.9\textwidth}
  \centering
 \includegraphics[width=\textwidth]{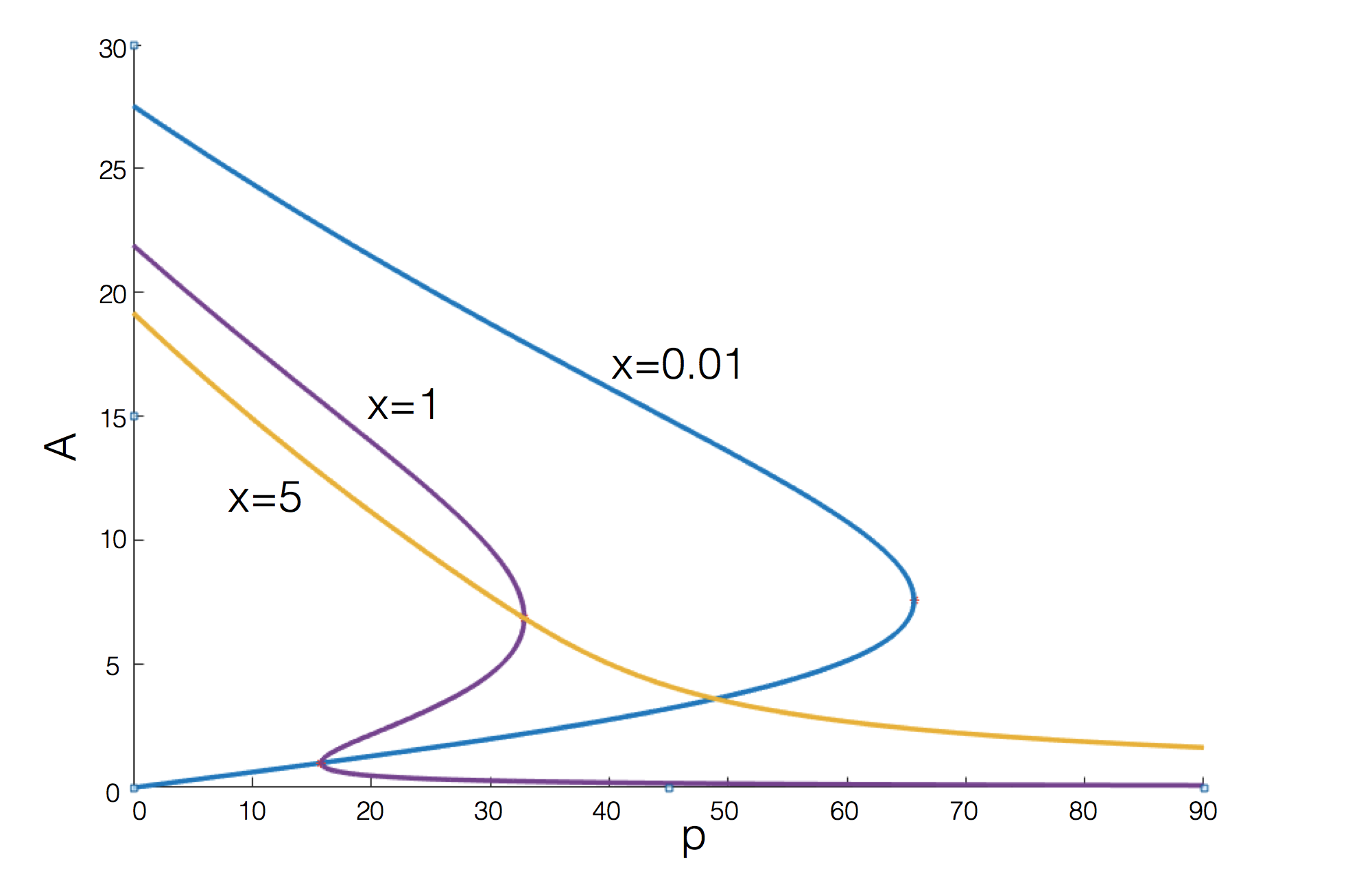}
 \caption{Positive equilibrium  adult densities with varying $p$ at different $x$ values. }
\end{subfigure}
\caption{Different types of adult positive steady states. $p$ represents the number of predators, and $x$ represents the density of prey at half of the maximum feeding rate.}
\label{fig:States}
\end{figure}

If we choose the Ricker recruitment function (\texttt{rickerm}) and linear fishing function, then model \eqref{3} becomes
\begin{equation}\label{5}
\begin{split}
\frac{dJ}{dt} &= \frac{\alpha A}{1+bA^2} -(p+A) \frac{k_{max}J^2}{x^2+J^2}-\beta J,\\
\frac{dA}{dt} &= \beta J-mA - f_l A.
\end{split}
\end{equation} 
Using estimated parameters listed in Table \ref{tab:Parameters} except $x$ and $p$, we vary $x$ and $p$ to explore the number of positive equilibria (Figure \ref{fig:States}(a)). When $x$ is large, there is only one positive equilibrium, as shown by the yellow curve $x=5$ in Figure \ref{fig:States}(b). The only positive equilibrium is stable, and $A=0$ (extinction) is always an unstable equilibrium as here $m+f_l=1.194<30.62=\alpha$. When $x$ is relatively small, we can detect multiple positive equilibria, such as the purple curve when $x=1$ (Figure \ref{fig:States}(b)). This means that when $p$ is small, there is a high stable equilibrium. Then, as $p$ grows larger, there will be a high and a low positive equilibrium, with an unstable state in between, whereas the extinction state is unstable. At a higher $p$, only the low stable positive equilibrium exists. However, when $x$ is very small, there will be only two positive equilibria at smaller $p$ (e.g., blue curve in Figure \ref{fig:States}(b). The higher one is stable, and the lower one is unstable. When $p$ gets too large, there will be no positive equilibria. 

While having 0, 1, 2 or 3 positive equilibria is mathematically possible, they are not likely to happen with biologically realistic $x$ values. The range of $x$ is most likely to be higher than 3 since prey-switching should occur within the observed range of $J$ values, as discussed previously. This indicates that there will only be one positive stable state and an unstable zero state (extinction). 

The model with a hyperbolic fishing term was analyzed similarly. Although the linear fishing mortality eventually exceeded hyperbolic fishing at high adult densities, there was little difference in the bifurcation diagrams. One possible explanation is that adult female density at equilibrium is low, so the linear and hyperbolic fishing terms behave similarly.

\subsection*{Comparison between linear and sigmoidal predation/cannibalism}

\begin{figure}[ht!]
\begin{subfigure}{0.8\textwidth}
  \centering
  \includegraphics[width=\textwidth]{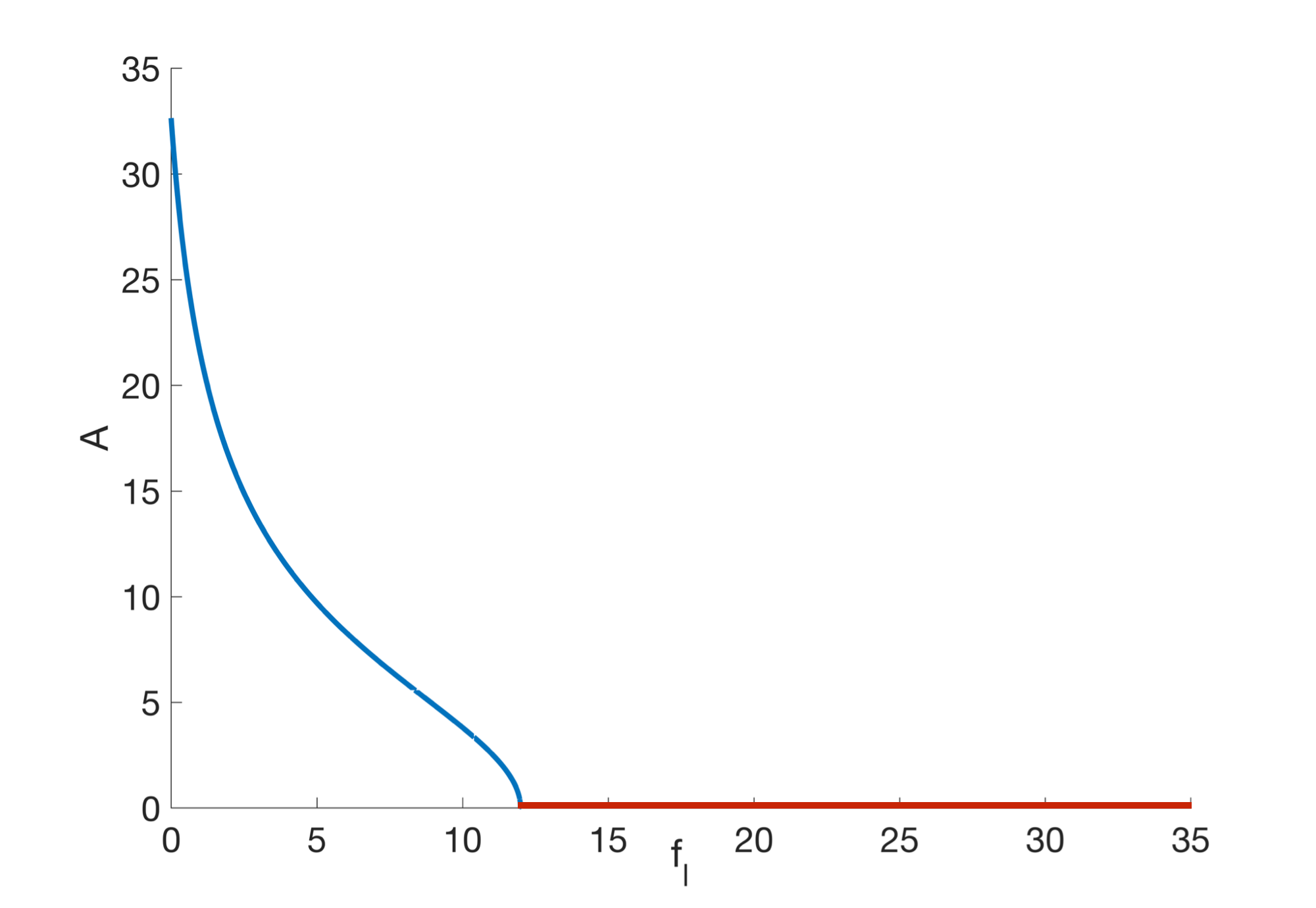}  
  \caption{Stable adult density for varying $f_l$ (linear fishing rate) with constant juvenile predation and cannibalism rate.}
\end{subfigure}
\begin{subfigure}{0.8\textwidth}
  \centering
  \includegraphics[width=\textwidth]{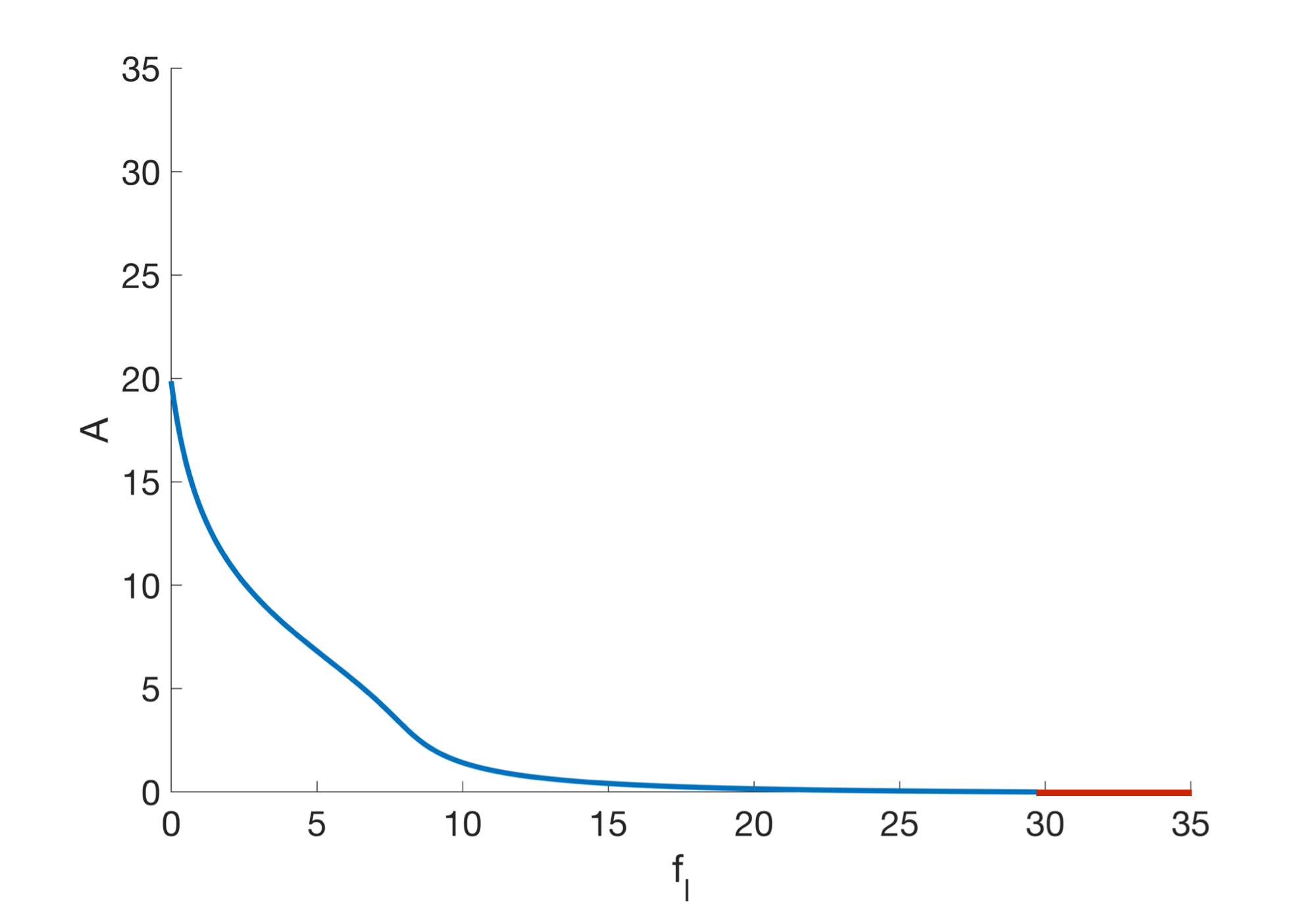}  
  \caption{Stable adult density for varying $f_l$ (linear fishing rate) with density-dependent juvenile predation and cannibalism rate.}
\end{subfigure}
\caption{Comparison of stable adult density under constant and density-dependent predation and cannibalism. Blue curve: positive stable state; red line: stable extinction state ($A=0$) at high $f_l$.}
\label{fig:StableState}
\end{figure}

In the simplest case, predation and cannibalism of juvenile crabs and fishing of adult crabs are assumed to be linear functions. The model is in the form
\begin{equation}\label{6}
\begin{split}
\frac{dJ}{dt} &= \frac{\alpha A}{1+bA^2} -1.5J-\beta J,\\
\frac{dA}{dt} &= \beta J-mA - f_l A.
\end{split}
\end{equation} 
In this case, the positive equilibrium $(J,A)$ of \eqref{6} is unique and it can be explicitly solved:
\begin{equation*}
    A=\sqrt{\left(\frac{\alpha \beta}{(1.5+\beta)(m+f_l)}-1\right)/b}, \; J=(m+f_l)A.
\end{equation*}
This positive equilibrium only exists if the fishing rate $f_l<\ds\frac{\alpha \beta}{(1.5+\beta)}-m=f_l^*=11.99$ (using the parameter values in Table \ref{tab:Parameters}). Figure \ref{fig:StableState}(a) shows the stable adult density for \eqref{6} for varying linear fishing rates. 

When the linear fishing rate $f_l$ is below $f_l^*$, there exists a positive stable equilibrium, and the extinction state ($A=0$) is unstable. When $f_l$ is larger than $f_l^*$, there is no positive equilibrium, and the extinction state is the only stable state, which represents extreme overfishing.

Figure \ref{fig:StableState}(b) shows the stable adult density for varying $f_l$ with sigmoidal predation and cannibalism (same as \eqref{5}). Similar to \ref{fig:StableState}(a), we have a stable positive equilibrium and an unstable zero when $f_l$ is low. When $f_l$ is larger, we have the extinction state as the only stable equilibrium. But in this case, the critical fishing rate is $f_l^*=\alpha-m=29.72$.

Compared with linear predation and cannibalism, blue crabs can survive a higher fishing rate with sigmoidal predation and cannibalism because of the feature that mortality rate under sigmoid predation and cannibalism declines at low prey densities. In contrast, the shape of the curve in Figure \ref{fig:StableState}(b) for the model with sigmoidal predation and cannibalism follows the one in Figure \ref{fig:StableState}(a) with constant predation and cannibalism when $f_l$ is high. But when the fishing rate is low, adult density under sigmoidal predation and cannibalism is lower than that under constant predation and cannibalism.

\section*{Conclusions}

Based on simulations with our stage-structured demographic model, we believe that for the blue crab population in Chesapeake Bay: (1) bistable positive states are not likely with biologically realistic parameter values; (2) hyperbolic (depensatory) fishing did not produce extinction at the range of juvenile and adult densities observed in the WDS; and (3) crabs can survive a higher fishing rate under the more realistic assumption of sigmoidal (density-dependent) predation and cannibalism than under constant (density-independent) predation and cannibalism. Our model could produce bistability under unrealistic parameter values. While we are confident about the estimation of most terms in the model, the parameters in the predation and cannibalism term had a wide range of possible values, indicating the need for further study of these functions. By comparing the bifurcation diagram of linear fishing rate with different forms of predation and cannibalism, we observed that under sigmoidal predation and cannibalism, blue crabs could survive higher fishing rate, suggesting that the blue crab population in Chesapeake Bay is more resistant to fishing than would be expected from density-independent predation and cannibalism \citep{Miller2011,MDNR2019}.

\bibliographystyle{apalike}

\newpage

\begin{table}[h]
\caption{Parameters in the model.}\label{tab:Parameters}
\bigskip
\centering
\begin{tabular}{cccc}
  \hline \hline
Parameter & Description & Units & Value \\
\hline
$\alpha$ & Maximum per capita reproduction rate & $y^{-1}$& 30.62 \\
$b$ & Density-dependent effect on reproduction & number$^{-2}$ $\cdot 10^{6} m^{4}$ & 0.01256 \\
$p$ & Predator density &number $\cdot 10^{-3}m^{-2}$& 
\\
$k_{max}$ &Maximum feeding rate&$y^{-1}$&  \\
$x$ & Prey density at $\frac{1}{2}$ maximum feeding rate&number $\cdot 10^{-3}m^{-2}$&  \\
$\beta$ & Maturation rate of juveniles&$y^{-1}$ & 1.09 \\
$m$ & Adult mortality rate&$y^{-1}$ & 0.9 \\
$f_l$ & Linear fishing mortality rate & $y^{-1}$& 0.294 \\
$f_h$ & Maximum fishing mortality rate & $y^{-1}$ & 37.097 \\
$s$ & Adult density at $\frac{1}{2}$ maximum fishing rate & number $\cdot 10^{-3}m^{-2}$  & 69.130 \\
\hline

\end{tabular}
\end{table}

\newpage

\begin{table}[h]
\caption{Different forms of the recruitment term $R(A)$, where $\alpha$ and $b$ are fit parameters.}\label{tab:Recruitment}
\bigskip
\centering
\begin{tabular}{ccc}
  \hline \hline
Model &  Description & Equation\\
\hline
\texttt{null} & Constant & $\alpha$ \\ \\
\texttt{linearm} & Linear & $\alpha A$ \\ \\
\texttt{hyperm} & Hyperbolic & $\ds\frac{\alpha A}{1+bA}$\\ \\
\texttt{hyperAm} & Hyperbolic with Allee effect  & $\ds\frac{\alpha A^2}{1+bA^2}$\\ \\
\texttt{rickerm} & Ricker & $\ds\frac{\alpha A}{1+bA^2}$   \\ \\
\texttt{rickerAm} & Ricker with Allee effect &  $\ds\frac{\alpha A^2}{1+bA^3}$   \\ \\
\hline
\end{tabular}
\end{table}

\newpage

\begin{table}[h]
\caption{Goodness of fit results of different models for the recruitment term ranking from the lowest to the highest AICc. K: number of parameters including variance; RSS: residual sum of squares; AICc: Akaike Information Criterion with small sample adjustment; Weight: probability of the model being the best among all models.}\label{tab:AICc} 
\bigskip
\centering
\begin{tabular}{ccccc}
  \hline \hline
Model &  K & RSS & AICc & Weight\\
\hline
\texttt{hyperAm} & 3& 55310.5 & 298.9 & 0.27 \\
\texttt{null} & 2 & 60854.4 & 299.1 & 0.25\\
\texttt{hyperm} & 3 & 55650.0 & 299.1 & 0.25\\
\texttt{rickerm} & 3 & 56729.0 & 299.6 & 0.19 \\
\texttt{rickerAm} & 3 & 62503.8 & 302.4 & 0.05\\
\texttt{linearm} &2 &102499.5 & 313.7& $<$0.01\\
\hline

\end{tabular}
\end{table}

\newpage
\section*{Figures}

\newpage
 
\newpage
 
\newpage
  
 \newpage

\newpage

\newpage

\newpage

\newpage

\end{spacing}

\end{document}